\documentclass{aa}  

\usepackage{graphicx}
\usepackage{txfonts}
\usepackage{natbib}
\bibpunct{(}{)}{;}{a}{}{,}   
\usepackage{hyperref}
\def\kms{km\,s$^{-1}$}
\begin{document}
  
   \title{A study of rotating globular clusters - the case of the old, metal-poor globular cluster NGC~4372
  \thanks{Based on observations made with ESO telescopes at the La Silla Paranal Observatory under programmes ID 088.B-0492(A), 088.D-0026(D), 164.O-0561, 71.D-02191B.}
  }
   \titlerunning{NGC~4372}
   
    \author{N. Kacharov\inst{1}\thanks{Member of the International Max Planck Research School for Astronomy and Cosmic Physics at the University of Heidelberg, IMPRS-HD, Germany.}
          \and
            P.~Bianchini\inst{2}$^{\star\star}$
          \and
            A.~Koch\inst{1}
          \and
            M.~J.~Frank\inst{1}
          \and
            N.~F.~Martin\inst{3,2}
          \and
	    G. van de Ven\inst{2}
          \and
            T.~H.~Puzia\inst{4}
          \and
            I. McDonald\inst{5}
          \and
            C.~I.~Johnson\inst{6}
          \and
            A.~A.~Zijlstra\inst{5}
          }

   \institute{Landessternwarte, Zentrum f\"{u}r Astronomie der Universit\"{a}t Heidelberg, K\"{o}nigstuhl 12, D-69117 Heidelberg, Germany\\
              \email{n.kacharov@lsw.uni-heidelberg.de}
         \and
              Max Planck Institut f\"{u}r Astronomie, K\"{o}nigstuhl 17, D-69117 Heidelberg, Germany
         \and
              Observatoire astronomique de Strasbourg, Universit\'{e} de Strasbourg, CNRS, UMR 7550, 11 rue de l'Universit\'{e}, F-67000 Strasbourg, France
         \and
              Institute of Astrophysics, Pontifica Universidad Cat\'{o}lica de Chile, Avenida Vicu\~{n}a Mackenna 4860, Macul, Santiago, Chile
         \and
              Jodrell Bank Centre for Astrophysics, Alan Turing Building, Manchester M13 9PL, UK
         \and
              Harvard-Smithsonian Center for Astrophysics, MS-15, 60 Garden Street, Cambridge, MA 02138, USA
              }

   \date{Received 26 Feb. 2014 / Accepted 2 June 2014}


  \abstract
   {NGC~4372 is a poorly studied old, very metal-poor Globular Cluster (GC) located in the inner Milky Way halo.}
   {We present the first in-depth study of the kinematic properties and derive the structural parameters of NGC~4372 based on the fit of a Plummer profile and a rotating, physical model. We explore the link between internal rotation to different cluster properties and together with similar studies of more GCs, we put these in the context of globular cluster formation and evolution. }
   {We present radial velocities for 131 cluster member stars measured from high-resolution FLAMES/GIRAFFE observations. Their membership to the GC is additionally confirmed from precise metallicity estimates. Using this kinematic data set we build a velocity dispersion profile and a systemic rotation curve. Additionally, we obtain an elliptical number density profile of NGC~4372 based on optical images using a MCMC fitting algorithm.
   From this we derive the cluster's half-light radius and ellipticity as $r_h=3.44\arcmin\pm0.04\arcmin$ and $\epsilon=0.08\pm0.01$. Finally, we give a physical interpretation of the observed morphological and kinematic properties of this GC by fitting an axisymmetric, differentially rotating, dynamical model.}
   {Our results show that NGC~4372 has an unusually high ratio of rotation amplitude to velocity dispersion ($1.2$ vs. $4.5$~km\,s$^{-1}$) for its metallicity. This, however, puts it in line with two other exceptional, very metal-poor GCs - M~15 and NGC~4590. We also find a mild flattening of NGC 4372 in the direction of its rotation. Given its old age, this suggests that the flattening is indeed caused by the systemic rotation rather than tidal interactions with the Galaxy. Additionally, we estimate the dynamical mass of the GC $M_{dyn}=2.0\pm0.5\times10^5~\mathrm{M_{\odot}}$ based on the dynamical model, which constrains the mass-to-light ratio of NGC~4372 between $1.4$ and $2.3~\mathrm{M_{\odot}/L_{\odot}}$, representative of an old, purely stellar population.}
   {}

   \keywords{Globular clusters: general --
             Globular clusters: individual: NGC~4372 --
             Galaxy: halo --
             Galaxy: kinematics and dynamics
               }

   \maketitle
%

\section{Introduction}


For a long time Globular Clusters (GCs) have been viewed as spherically symmetric, non-rotating stellar systems, successfully described to first order \citep[see][]{trager+95,mclaughlin+vdmarel05} by spherical, isotropic models \citep[e.g.][]{king66,wilson75}.
However, the increasing abundance of observational data revealed noticeable deviations from this simple picture.
Indeed, radial anisotropy \citep{ibata+2013}, significant degree of mass segregation \citep{dacosta82}, signatures of core-collapse \citep{newell+oneil78,djorgovski+king84}, velocity dispersion inflated by binaries \citep{bradford+2011} and mild deviation from sphericity often associated with the presence of tidal tails \citep{white+shawl87,odenkirchen+2001,belokurov+2006,chen+chen2010} have been observed in these stellar systems.
Additionally, significant amounts of internal rotation \citep{lane+2011,bellazzini+2012,bianchini+2013} have been observed in most Milky Way GCs. This has motivated the development of dynamical models including the effects of external tides, mass segregation, core-collapse and binary stars \citep{gunn+griffin79,kuepper+2010,zocchi+2012}, as well as a significant degree of rotation \citep{wilson75,satoh80,davoust86,vandeven+2006,fiestas+2006,varri+bertin2012}.


NGC~4372 is a relatively nearby ($\mathrm{R_{\odot}}=5.8$~kpc), and yet neglected GC in the inner halo. Photometric studies have established it as an archetypical old ($> 12$~Gyr) and metal-poor object, [Fe/H]~$\simeq-2.1$~dex \citep{alcaino+91,geisler+95,rutledge+97,rosenberg+2000,piotto+2002}.
It is also of particular interest from a dynamical point of view, since it is known to harbour close stellar binaries and luminous X-ray sources \citep{servillat+2008} - curiously these are mainly found outside the central regions.
This is in contrast to the expected segregation based on the larger dynamical mass of such binary systems and suggests that NGC~4372 has still not established a significant degree of kinetic energy equipartition or that other dynamical processes have stirred up the cluster and expelled these sources from the centre.
With a core relaxation time $\log t_c = 8.88$~dex \citep{harris96}, NGC~4372 is an intermediately relaxed system, according to the classification of \citet{zocchi+2012}.
Considering its relatively low concentration and old age, it might be an example of a re-bounced, post core-collapse GC \citep[see][]{cohn+hut84}.
In this context, we also note its short orbital period ($0.1$~Gyr) and moderate vertical space velocity component ($W=+100$~\kms), so that the resulting orbit \citep{dinescu+2007} implies many slow disk crossings.
The similarities of the orbit of NGC~4372 to the orbit of the massive GC NGC~2808 has suggested that both clusters are dynamically paired.

Internal rotation is one of the main reasons for the flattening of GCs, but external tides and pressure anisotropy can also play a significant role \citep{vandenbergh2008}. In this work we build a dynamical, rotating model of the old, metal-poor, halo GC NGC~4372 using the \citet{varri+bertin2012} family of models \citep[see also][]{bianchini+2013}. From this, we derive and use the maximum rotation amplitude to central velocity dispersion ratio ($A_{rot}/\sigma_0$) to assess the importance of internal rotation in this GC.
While, it has been shown, by means of N-body simulations, that internal rotation and tidal interactions significantly accelerate the dynamical evolution of GCs \citep{boily2000,ernst+2007,kim+2008}, it is not yet clear how they affect the formation and the earliest stages of GC evolution.
In an attempt to shed some light on this question, we investigate the impact of internal rotation on various cluster parameters like horizontal branch morphology, age, metallicity, and chemical variations.


This paper is organised as follows. Section 2 describes the observations and the data reduction. Section 3 is dedicated to the derivation of the structural parameters of NGC~4372 from photometry. Section 4 is dedicated to a detailed kinematic study of this object. Our discussion is developed in Section 5, where we present the rotating model and compare the kinematic results of NGC~4372 with other Milky Way GCs with existing kinematic data.

\section{Observations and data reduction}

\subsection{Spectroscopy}

\begin{figure}
   \centering
   \includegraphics[width=\hsize]{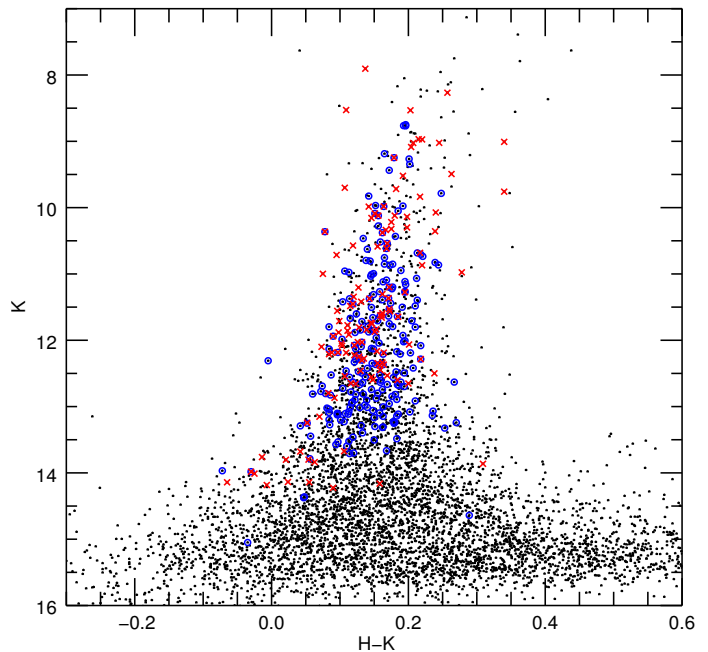}
      \caption{2MASS CMD of NGC~4372. The blue circles indicate all stars (both cluster members and foreground) in the first (088.B-0492) and red crosses indicate all stars in the second (088.D-0026) spectroscopic samples.
              }
         \label{fig:CMD}
   \end{figure}

The targets were selected from archival FORS2 pre-imaging in the B and V filters (ESO-programme 71.D-02191B, P.I.: L.~Rizzi) and the 2MASS catalogue \citep{cutri+03} to cover the entire span of NGC~4372's red giant branch (RGB) and includes a number of asymptotic giant branch (AGB) stars (Figure \ref{fig:CMD}). The spectroscopic observations were carried out in service mode in the nights of Feb. 11, Mar. 08, and Mar. 10, 2012 using the Fibre Large Array Multi Element Spectrograph (FLAMES) mounted at the UT2 (Kueyen) of the Very Large Telescope (VLT) on Paranal \citep{pasquini+2002}.
Five observing blocks (OB) were executed in total (exposure time $2775$~s per OB) using two different Medusa plates. On each mask $133$ fibres were fed to the GIRAFFE spectrograph (using the HR13 grating, which covers the wavelength range $6100 - 6400$~\AA~ with spectral resolution R~$\sim 22000$) and 8 fibres were fed to the UVES spectrograph.
Nineteen of the GIRAFFE fibres were dedicated to the sky and $112$ to the RGB/AGB targets. Both plate settings include different targets with a large overlap between them.

The FLAMES observations were reduced with the standard GIRAFFE pipeline, version 2.9.2 \citep{Blecha+2000}. This pipeline provides bias subtraction, flat fielding, and accurate wavelength calibration from a Th-Ar lamp. The $19$ sky spectra were combined and subtracted from the object spectra with the IRAF task $skysub$. We computed radial velocities of our targets by crosscorrelating the spectra with a synthetic RGB spectrum with similar stellar parameters, as expected for our targets, using the IRAF $fxcor$ task.
All spectra were Doppler-shifted to the heliocentric rest frame and the individual spectra of the same stars (ranging from 2 to 5) were median combined using the IRAF $scombine$ task. Finally, the spectra were normalised to the continuum level with the help of the IRAF $continuum$ task.
The final, reduced, one-dimensional spectra have average signal-to-noise ratios (SNR) ranging from $20$ to $200$ per pixel, depending on the brightness of the stars and the number of the combined individual exposures.
The data set consists of $108$ different stars with successfully measured radial velocities, of which $64$ were confirmed cluster members and the rest were identified as foreground stars.

In the following analysis we also use a second GIRAFFE data set (ESO-programme 088.D-0026(D), P.I.: I.~McDonald) of $123$ stars observed in service mode in the nights of Jan. 15, Mar. 04, and Mar. 06, 2012 with the HR13 and HR14 GIRAFFE gratings, and thus, covering a total wavelength range from $6100$~\AA~to $6700$~\AA. It is reduced with the girBLDRS\footnote{\url{http://girbldrs.sourceforge.net/}} software.
The spectra cover predominantly the brighter RGB/AGB of NGC~4372 (Figure \ref{fig:CMD}) and are of high quality (median SNR of $\sim100$~per pixel). A radial velocity check and a metallicity estimate confirms $74$ of the stars as cluster members, while the rest are classified as foreground contaminants. There is very little overlap between the two samples and we found only eight stars in common, five of which are cluster members.
The radial velocities of the common stars agree to within $1$~\kms (mean difference $0.11$~\kms with $1$~\kms standard deviations).

For all stars in both data sets we also have metallicity estimates, which will be presented in a subsequent work. The membership of each cluster star is established based on simultaneous radial velocity and metallicity cuts.

\begin{figure}
   \centering
   \includegraphics[width=\hsize]{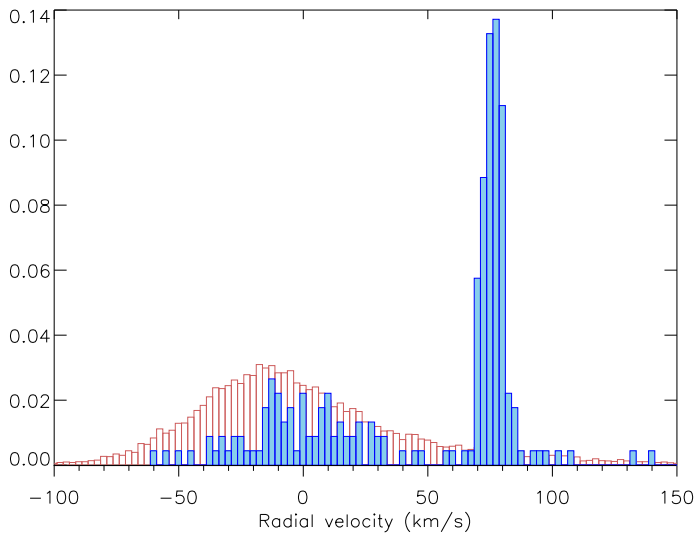}
      \caption{Radial velocities relative frequency of both spectroscopic samples (filled, blue boxes) overplotted over a Besan\c{c}on model of the field stars in the direction of NGC~4372 (blank, red boxes).
              }
         \label{fig:besancon}
   \end{figure}

Both data sets have essentially the same mean velocity and velocity dispersion, so we decided that it is safe to combine them.
Our final NGC~4372 spectroscopic sample consists of $131$ unique cluster member stars with K-band magnitude $\lesssim15$~mag, confirmed from metallicity estimates ($\mathrm{[Fe/H]}<-2.0$) of all the stars in the sample that have radial velocities between $50$ and $100$~\kms.
The median accuracy of the velocity measurements is $1$~\kms. In Figure \ref{fig:besancon} we compare the radial velocities distribution of our full spectroscopic sample (cluster members plus foreground contamination) with the velocity distribution in this direction of the sky according to the Besan\c{c}on model of the Galaxy \citep{robin+03}.
The Besan\c{c}on model predicts about $2100$ stars in an area of $0.2~\mathrm{deg^2}$ within our colour-magnitude selection limits.
The mean radial velocity of NGC~4372 of $76$~\kms~is, however, quite distinct from the radial velocities of the majority of foreground stars, although a small contamination by such foreground stars is not excluded.
But considering the very low metallicity of this GC ([Fe/H]~$\simeq-2.2$~dex) and the clearly metal rich foreground stellar population, we are confident that we have selected a clean sample of NGC~4372 member stars -- about $100$ stars from the adopted Galactic model have radial velocities between $50$ and $100$~\kms and none of them is more metal poor than $-1.8$~dex.
We come back to the detailed kinematics of these GC stars in Section 4.

\subsection{Photometry}


We used archival imaging obtained with the Wide Field Imager (WFI) at the 2.2m MPG/ESO telescope at La Silla \citep{baade+99}. We chose $V$- and $I$-band observations of the cluster taken as part of the `pre-FLAMES' programme of the ESO Imaging Survey (programme 164.O-0561, PI: Krautter; cf. \citealt{momany+2001}), which cover a field of view (FOV) $30' \times 30'$ centred on the cluster.
The basic data reduction, astrometric solution, and combining of mosaics were carried out using the \textsc{theli} pipeline \citep{erben+2005,schirmer2013}. We used the Two Micron All Sky Survey (2MASS) point source catalog \citep{skrutskie+2006} as astrometric reference and combined the observations to one stacked, undistorted image per filter.
From these images we then obtained instrumental magnitudes for unresolved objects using the \textsc{daophot} software package \citep{stetson87,stetson93}. Photometric zero points were fixed to standard stars in the same field from the standard star database of \citet{stetson2000,stetson2005}\footnote{Available at \url{http://www3.cadc-ccda.hia-iha.nrc-cnrc.gc.ca/community/STETSON/standards/}}. Since we are primarily interested in relative photometry rather than absolute photometry in a given standard system, no colour term was included.

\begin{figure*}
   \centering
   \includegraphics[width=13.5cm]{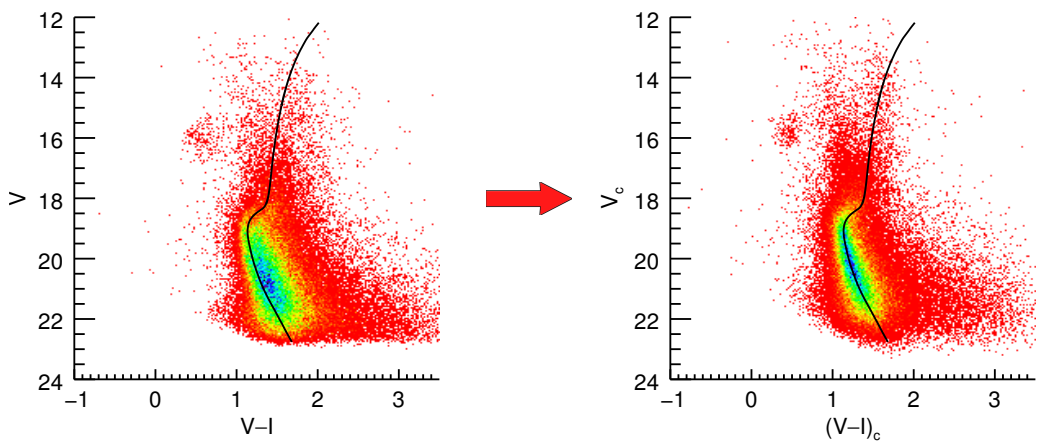}
      \caption{Hess diagrams of NGC~4372's CMD before ({\it left panel}) and after ({\it right panel}) the differential reddening correction. The BASTI isochrone used as a fiducial line is overimposed.
              }
         \label{fig:dereddening}
   \end{figure*}

\begin{figure}
   \centering
   \includegraphics[width=\hsize]{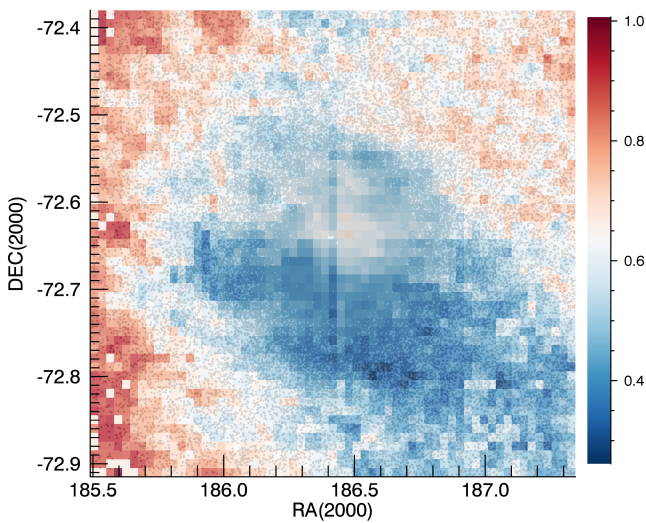}
      \caption{A reddening map across the field of NGC~4372.
              }
         \label{fig:reddening_map}
   \end{figure}

Since NGC~4372 resides behind a strip of Galactic gas and dust, it suffers from severe differential reddening \citep{hartwick+73,alcaino+91,gerashchenko+99}. In an attempt to correct for this effect we followed the procedures described in \citet{hendricks+2012} and \citet{milone+2012}.
The main idea is to estimate the interstellar extinction for each star individually based on the median distance of its nearest neighbours to a fiducial line along the reddening vector. In our case, we used an adaptive number of the nearest neighbours depending on the density of the region, starting from $40$ in the innermost $15\arcsec$ and using $10$ stars in the outermost regions. We used only main sequence stars that have uncertainties on the V- and I-band photometry less than $0.1$~mag.
A BASTI isochrone \citep{pietrinferni+04,pietrinferni+06} of old age ($15$~Gyr) and low metallicity ($Z=0.0003$), shifted to a distance modulus $(m-M)_V=15.0$~mag \citep{harris96}, was used as a fiducial line, representative for this GC.
Despite its relatively high age with respect to the age of the Universe, the chosen isochrone represents well the main sequence, the turn-off point, the RGB, and the magnitude of the HB.
We used the standard interstellar extinction law for the Milky Way where $A_V=3.1\times E(B-V) = 2.2\times E(V-I)$ \citep{cardelli+89,mathis90}.
The results of the dereddening procedure are presented in Figure \ref{fig:dereddening}, where we show a Hess diagram of the original raw colour-magnitude diagram (CMD) and the resulting CMD after applying the described algorithm.
The dereddening procedure also corrects for residual variations in the photometry that result from the variation of the point spread function or illumination \citep[e.g.][]{koch+2004} across the combined mosaics.
We find a mean $E(B-V)\sim0.5$~mag towards NGC~4372 in a good agreement with the \citet{schlafly+finkbeiner2011} recalibration of the \citet{schlegel+98} extinction map, with a significant variation between $0.3$ and $0.8$~mag across the field of the GC. A reddening map is presented in Figure \ref{fig:reddening_map}.

\begin{figure}
   \centering
   \includegraphics[width=\hsize]{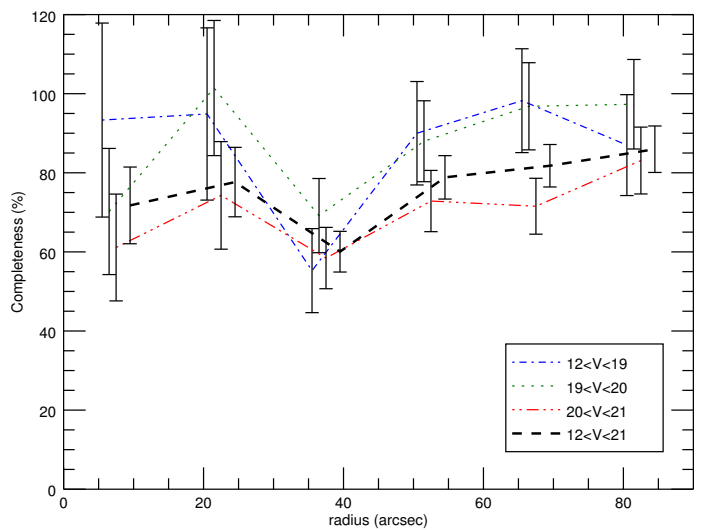}
      \caption{Completeness of the WFI photometry compared to Hubble imaging of the innermost $1.5\arcmin$ of NGC~4372 in different magnitude and spatial bins. The radial bins are shuffled by $1\arcsec$ for clarity.
              }
         \label{fig:completeness}
   \end{figure}

To assess the completeness of the photometry, the final WFI photometric catalogue was compared to HST imaging of the central regions of the cluster \citep{piotto+2002} that covers an area with radius of about $1.5\arcmin$ and can be safely assumed to be complete down to $V=21$~mag.
The HST F555W band magnitudes were calibrated to the WFI $V$-band extinction corrected magnitudes of matching targets. We assumed an uniform extinction across the HST field, which is justified by its small size.
We considered different magnitude ranges and counted the stars in both catalogues in $6$ concentric rings each $15\arcsec$ thick assuming Poisson uncertainties. The results are presented in Figure \ref{fig:completeness}.
Considering all stars, the comparison shows a completeness level of $(70\pm10)\%$ in the innermost $15\arcsec$ that rises to about $(85\pm5)\%$ at $90\arcsec$ from the cluster centre. This incompleteness is mostly due to the faintest stars in the sample with $V>20$~mag, where the central $15\arcsec$ region has a completeness level of $(60\pm15)\%$ that rises to $(80\pm10)\%$ in the outer radii. The stars brighter than $V=20$~mag have a completeness level of about $(90\pm15)\%$ across the whole field. An exception is the region at $\sim30\arcsec$, where we observe a sudden drop of completeness in the whole magnitude range.

\section{Structural parameters}

According to the Harris catalogue of GCs \citep[][2010 version\footnote{\url{http://physwww.mcmaster.ca/~harris/mwgc.dat}}]{harris96}, NGC~4372 has a half-light radius $r_h = 3.91\arcmin$, a tidal radius $r_t = 34.9\arcmin$, and a concentration parameter $c = \log(r_t/r_c) = 1.3$, where $r_c = 1.75\arcmin$ is its core radius. These estimates come from a poorly constrained surface brightness profile presented in \citet{trager+95}. Furthermore, there is discrepant information for the ellipticity of NGC~4372 in the literature \citep{white+shawl87,chen+chen2010}. Therefore, we decided to independently re-derive these parameters.
It is interesting to check the direction of the flattening to assess the importance of internal rotation (see Section 4.2), or give some insight on its tidal interaction with the Galaxy.

Using the WFI photometric catalogue, corrected for differential reddening and incompleteness in the central regions, we construct a number density, instead of a surface brightness profile, since the former is less sensitive to the presence of individual bright stars and foreground contaminants. It also allows us to work with the full catalogue, instead with a limited number of surface bins.
We considered only stars brighter than $V=20$~mag, where we can safely assume that the sample is $100\%$ complete beyond $90\arcsec$.

\subsection{The method}

We followed a modified version of the maximum likelihood method outlined in \citet{martin+2008} to fit the GC star number density profile. This approach consists of maximizing the log-likelihood function (Eq. 1) by iterating a set of free parameters, for which the observations become most probable:
\begin{equation}
 \log\mathcal{L}(p_1,p_2,...,p_j) = \sum_i \log\ell_i(p_1,p_2,...,p_j)
\end{equation}
where $\ell_i(p_1,p_2,...,p_j)$ is the probability of finding the measurement $i$ given the set of parameters $p_1,p_2,...,p_j$. We used a set of $6$ free parameters for our fit - the centroid of the cluster $(x_0, y_0)$, a model-dependent characteristic radius, the global ellipticity\footnote{The ellipticity is defined as $\epsilon=1-\frac{b}{a}$, where $a$ and $b$ correspond to the semi-major and semi-minor axis of the ellipse, respectively.} ($\epsilon$) with the orientation angle from north to east ($\theta$) of the major axis, and the contaminating foreground density ($n_f$) in stars per square arcmin. The centroid is defined as in \citet{vandeven+2006}:
\begin{equation}
 x_i - x_0 = \sin(\alpha_i-\alpha_0)\cos\delta_i
\end{equation}
\begin{equation}
 y_i - y_0 = \sin\delta_i\cos\delta_0 - \cos\delta_i\sin\delta_0\cos(\alpha_i - \alpha_0)
\end{equation}
where $\alpha_i$ and $\delta_i$ are the equatorial coordinates of the stars from the WFI catalogue and $\alpha_0=12^h 25^m 45.40^s$ and $\delta_0=-72^{\circ} 39\arcmin 32.4\arcsec$ are the central cluster coordinates listed in the Harris catalogue.
We chose to work with the projected Plummer family of models \citep{plummer1911} of the following type:
\begin{equation}
 n(r) = n_0 \left(1+\frac{r^2}{a^2}\right)^{-2} + n_f
\end{equation}
In the Plummer profile, the characteristic radius $a$ corresponds to the half-light radius\footnote{Technically, this is the radius that contains half of the stars. In order to relate this value to the half-light radius, one has to assume an initial mass function, binarity fraction, and segregation distribution. In this work we use this quantity as a reasonable approximation to the half-light radius.} of the cluster.
In the above equations, the independent variable $r$ is an elliptical radius (the semi-major axis of the adopted ellipse), which is related to the spatial position $(x,y)$ of the stars in the following way:
\begin{equation}
 r = \left\{\left[\frac{1}{1-\epsilon}(x\cos\theta-y\sin\theta)\right]^2 + (x\sin\theta+y\cos\theta)^2\right\}^{1/2}
\end{equation}
and $n_0$ is the central number density of the cluster. The central number density is not a free parameter but constrained from the total number of stars in the FOV within the selection criteria ($N_{tot}$). 
\begin{equation}
 N_{tot} = \int_{FOV} n(r) dx dy = n_0\int_{FOV}\left(1+\frac{r^2}{a^2}\right)^{-2}dxdy + A n_f
\end{equation}
where $A$ is the total area of the FOV. From the last expression we can write that
\begin{equation}
 n_0 = \frac{N_{tot}-An_f}{\int_{FOV}\left(1+\frac{r^2}{a^2}\right)^{-2}dxdy}
\end{equation}
The integration from the above equation is done numerically over the entire FOV by dividing it into small segments with sizes $\Delta x, \Delta y$ much smaller than the expected half-light radius of the cluster (few arcmin).


We iterated the parameters in a Markov Chain Monte Carlo (MCMC) manner following the Metropolis-Hastings algorithm \citep{hastings70}, where each new set of parameters is derived randomly from the previous set, from a Gaussian distribution function with a defined standard deviation. The standard deviation for each parameter is chosen such as to optimize the acceptance rate of the Markov chain.
A probability based on the likelihood is assigned to the new set of parameters and the chain is continued until we achieve a good sampling of the parameter space.
The code was extensively tested with Monte-Carlo drawn Plummer clusters with various half-light radii, ellipticities, and uniform back-ground densities to confirm the correctness of the results. \citet{munoz+2012} explored the conditions under which the outlined approach gives reliable results, namely the size of the FOV, the total number of stars, and the $n_0/n_f$ ratio, which in the case of NGC~4372 are all fulfilled.

\subsection{Profile fitting}

In order to obtain the global structural parameters of NGC~4372, we performed two fits, using all stars brighter than $V = 20$~mag and $V = 19$~mag, respectively. In both cases we assumed that the photometry is complete at radial distances greater than $1.5\arcmin$ but we applied completeness corrections according to Figure \ref{fig:completeness} in the central regions.

The results from the MCMC fits are summarized in Table \ref{tab:mcmc_fit}.
The uncertainty intervals in Table \ref{tab:mcmc_fit} are defined as the $1\sigma$ deviation from the mean of a Gaussian representing the distribution function of each parameter from the Markov chain after excluding the ``burn-in`` iterations.

\begin{figure}
   \centering
   \includegraphics[width=\hsize]{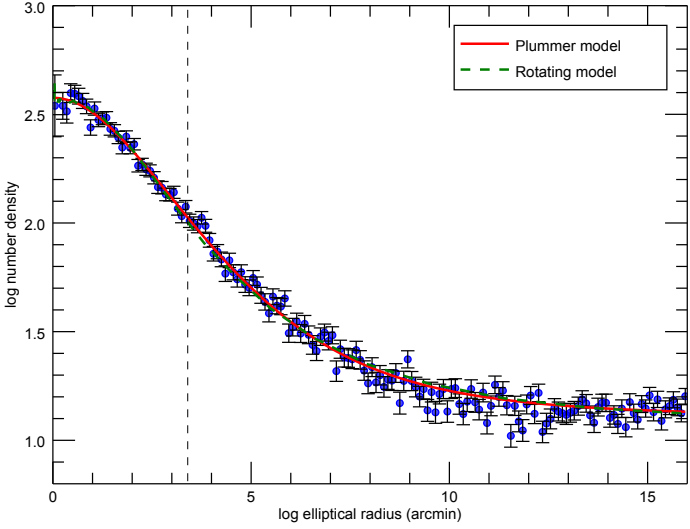}
      \caption{Incompleteness corrected number density profile of NGC~4372 built in elliptical annuli of size $0.1\arcmin$ (blue symbols) using stars brighter than $V=20$~mag. The best fitting Plummer model (red line), as well as the obtained rotating model (green line, see Section 5.1) are overimposed The fitted half-light radius is indicated with a vertical, dashed line.
              }
         \label{fig:number_density}
   \end{figure}

\begin{figure}
   \centering
   \includegraphics[width=\hsize]{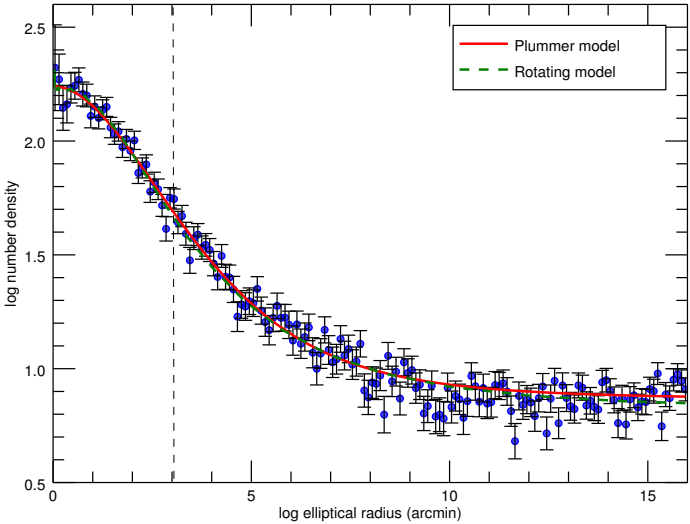}
      \caption{The same as Figure \ref{fig:number_density} but using only stars brighter than $V=19$~mag (the turn-off and RGB/AGB regions). 
              }
         \label{fig:number_density_bright}
   \end{figure}

Figures \ref{fig:number_density} and \ref{fig:number_density_bright} present the number density profiles of NGC~4372 obtained by using all stars brighter than $V=20$~mag and $V=19$~mag, respectively, built in confocal elliptical annuli with $0.1\arcmin$ size that have ellipticities and position angles in accordance to the results reported in Table \ref{tab:mcmc_fit} for both fits.
The best fitting Plummer profiles and rotating models (see Section 5.1) are also shown. We note here that the two curves drawn on the figures are not fits to the binned profiles but a result of the MCMC discrete fitting procedure in the case of the Plummer profile and the simultaneous modelling of the kinematics and number-density profiles in the case of the dynamical, rotating model.

When using stars within different magnitude ranges the derived half-light radii vary significantly (see Table \ref{tab:mcmc_fit}), while all other parameters used in the fit are consistent within the uncertainties.
According to the BASTI isochrones, the stellar mass in the considered magnitude range varies between $0.70$ and $0.75~M_{\odot}$.
Due to this small contrast of masses, together with the overall uncertainties associated with the dereddening procedure and completeness estimation across the large WFI field, it seems difficult to draw conclusions on the mass segregation of the cluster. Instead, we attribute the more extended distribution profile of the fainter stars to still unaccounted incompleteness in the photometry.
The very small field, however, in which incompleteness has been assessed for, does not allow us to properly account for it over the entire field of the WFI photometry.
Another possible source of uncertainty comes from the dereddening procedure, where we are using a particular isochrone to correct for differential extinction the GC stellar population. This isochrone is not representative for the field population, which might be partly over-corrected, thereby increasing the relative fraction of contaminating field stars in the fainter samples, making number density fits artificially broader.

\begin{table}
\begin{center}
\caption{Plummer model projected structural parameters from the MCMC fit.}\label{tab:mcmc_fit}
{\small
 \begin{tabular}{ccc}
\hline
 & $V<20$~mag & $V<19$~mag \\
\hline
$r_h$ (arcmin) & $3.44\pm0.04$  & $3.03\pm0.06$  \\
$x_0$ (arcmin) & $-0.33\pm0.03$ & $-0.33\pm0.05$  \\
$y_0$ (arcmin) & $-0.64\pm0.03$ & $-0.62\pm0.04$ \\
$\epsilon$     & $0.08\pm0.01$  & $0.09\pm0.02$  \\
$\theta$       & $48^{\circ}\pm6^{\circ}$ &  $52^{\circ}\pm6^{\circ}$  \\
$N_0$\tablefootmark{1}          & $12000\pm100$     & $4200\pm70$   \\
$n_f$ ($\star/\square$\arcmin)     & $12.82\pm0.16$ & $7.34\pm0.11$   \\
\hline
 \end{tabular}
\par}
\tablefoot{
\tablefoottext{1}{$N_0 = N_{tot} - A n_f$ is the estimated number of stars belonging to the cluster, given our magnitude cuts.}
}
\end{center}
\end{table}

In the following analysis we apply the derived centroid corrections ($x_0, y_0$) to the published coordinates of NGC~4372 in the Harris catalogue. The estimated coordinates of the centre are $\alpha_0=12^h 25^m 51^s$ and $\delta_0=-72^{\circ} 38\arcmin 57\arcsec$ for epoch 2000.

\section{Kinematics}

As mentioned in Section 2.1, we measured radial velocities from individual exposures using the $fxcor$ task in IRAF. This method applies a Fourier cross-correlation between the spectra of interest and a given template and provides velocity error estimates based on the fitted peak height and the antisymmetric noise \citep{tonry+davis79}. The adopted final velocities and their uncertainties are calculated as the error-weighted mean of the different velocity estimates for the same star from different exposures, ranging from 2 to 5 in our sample.
For the second sample we averaged the radial velocities measured in the HR13 and HR14 gratings for each star. The radial velocities derived from different exposures of the same star agree well with each other, showing that stellar binaries do not play a significant role.


We used a maximum likelihood approach \citep{walker+2006} to calculate the mean radial velocity of NGC~4372 ($v_r = 75.91 \pm 0.38$~\kms) and its global intrinsic velocity dispersion ($\sigma = 3.76 \pm 0.25$~\kms) from the discrete velocity measurements.
These values are in reasonable agreement with the radial velocity of $72.3\pm1.4$~\kms and velocity dispersion of $4.3\pm0.9$~\kms, estimated by \citet{geisler+95}, based on 11 stars.

\subsection{Central velocity dispersion}

\begin{figure}
   \centering
   \includegraphics[width=\hsize]{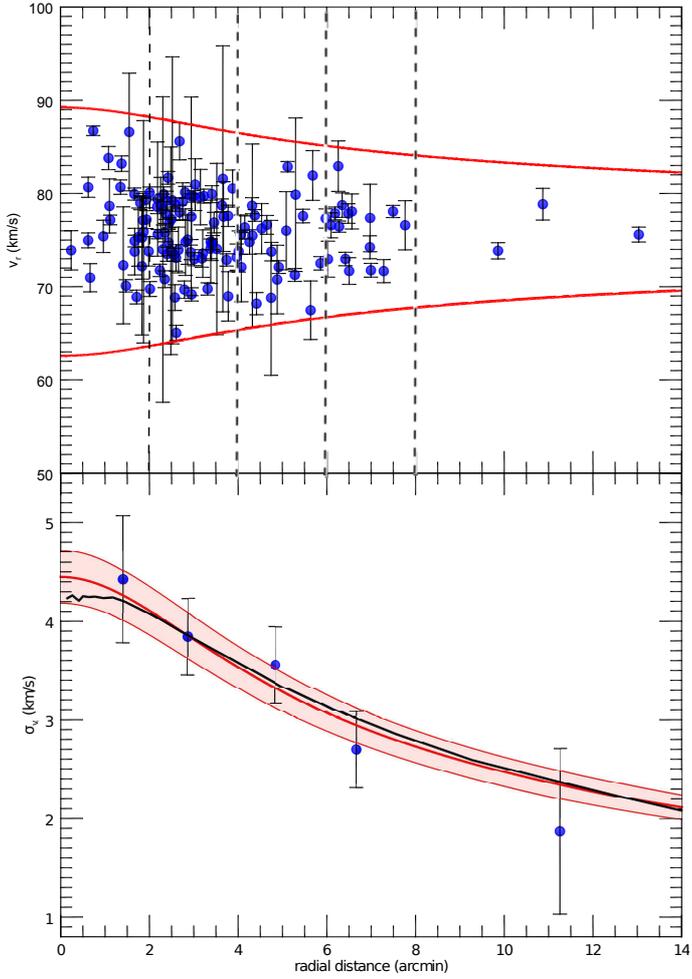}
      \caption{{\it Upper panel:} Radial velocities of the confirmed GC member stars versus radial distance from the centre of NGC~4372. The red curves indicate $\pm 3$ times the velocity dispersion as function of the radial distance. The vertical dashed lines indicate the borders of the bins used to compute the velocity dispersion at a given radius. {\it Bottom panel:} Velocity dispersion profile of NGC~4372 together with the best fitting Plummer profile (thick red line) and the rotating model (thick black line). The shaded area between the two thin red lines indicates the $1\sigma$ uncertainty of the Plummer profile.
              }
         \label{fig:plummer}
   \end{figure}

In order to estimate the central velocity dispersion $\sigma_0$ of NGC~4372, we divide the cluster into five radial bins and used the same maximum likelihood approach to estimate the velocity dispersion in each bin (Figure \ref{fig:plummer}), which we approximate with Plummer models.
Plummer models describe isotropic stellar systems with constant density cores. We do not claim that this family of models is the best representation for this particular GC but we rather use it as a reasonable approximation. Our velocity dispersion does not allow us to distinguish between different types of models, such as the more physical King models for example.
In principle, if mass follows light, as is expected to be the case for Galactic GCs \citep{lane+2010}, the characteristic radius $a$ should be equal to the half-light radius $r_h$ of the cluster.

We fit the resulting velocity dispersion profile in a least-squares sense with a projected \citet{plummer1911} model (Eq. 7) using a characteristic Plummer radius $a=3.03\arcmin$ estimated from the number-density profile for the brighter stars (RGB/AGB and turn-off stars; see Section 3.2) and setting the central velocity dispersion $\sigma_0$ as a free parameter.
\begin{equation}
 \sigma(r)^2 = \frac{\sigma_0^2}{\sqrt{1+\frac{r^2}{a^2}}}
\end{equation}

Although we have radial velocity measurements only for RGB/AGB stars in NGC~4372, using an estimate for the half-light radius value derived from the RGB/AGB and turn-off stars together is a compromise between having a statistically large sample of stars for constraining the number-density profile (see Figure \ref{fig:number_density_bright}) and using stars with a similar spatial distribution.

Our data suggests a best-fit $\sigma_0 = 4.56\pm0.3$~\kms. The uncertainty is the formal $1\sigma$ error computed from the covariance matrix.

Having estimated the central velocity dispersion, it is straightforward to derive a dynamical mass for NGC~4372. According to an isotropic Plummer model the mass is given by the following expression \citep[see e.g.][]{dejonghe87,mackey+2013}:
\begin{equation}
 M=\frac{64a\sigma_0^2}{3\pi G}
\end{equation}
Adopting a heliocentric distance of $5.8$~kpc (from the Harris catalogue), our radius translates to $a=5.1\pm0.05$~pc and we obtain $M=1.7\pm0.3\times10^5~\mathrm{M_{\odot}}$. This value is in agreement with the mass listed in \citet{mandushev+91}, $M = 1.3^{(+1.9)}_{(-0.8)}\times10^5~\mathrm{M_{\odot}}$ based on a mass-luminosity relation.
The absolute magnitude of NGC~4372 according to the Harris catalogue (2010 version) is $M_V=-7.79$~mag. This corresponds to a luminosity of $L_V/L_{\odot} = 1.1\times10^5$, from where we can estimate the mass-to-light ratio for this GC to be $M/L_V\sim1.5~\mathrm{M_{\odot}/L_{\odot}}$, typical of most GCs \citep{mandushev+91,pryor+meylan93}. Thus, NGC~4372 is a typical representative of the old, purely stellar populations without detectable amounts of dark matter.

\subsection{Rotation}

\begin{figure}
   \centering
   \includegraphics[width=\hsize]{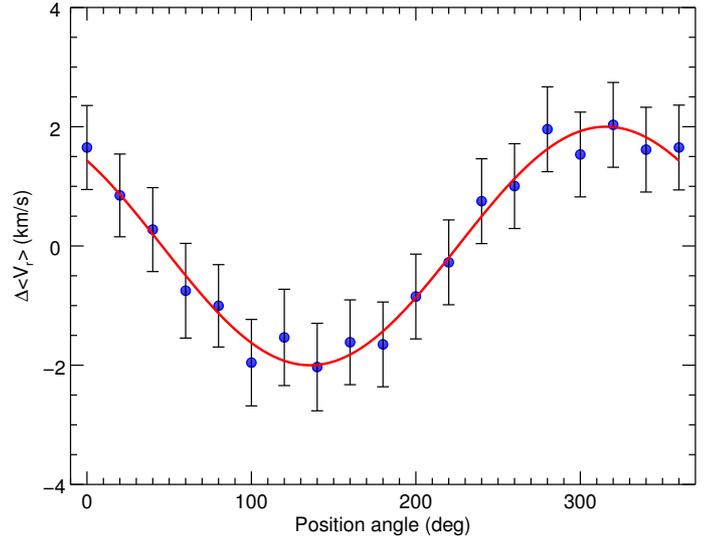}
      \caption{Rotation in NGC~4372. The plot displays the difference between the mean velocities on each side of the cluster with respect to a line passing through its centre at a given position angle (measured from north to east, north~$=0^{\circ}$, east~$=90^{\circ}$). The red line is the sine function that best fits the data.
              }
         \label{fig:rotation}
   \end{figure}

We checked for systemic rotation in NGC~4372, following a well established method \citep[see e.g.][and references therein]{mackey+2013,bellazzini+2012,lane+2009,lane+2010}: To this end we measured the difference between the maximum likelihood mean velocity on either side of a line passing through the cluster's centre and rotated at different position angles (Figure \ref{fig:rotation}).
The resulting curve is well described by a sine law of the type $\Delta v_r=A_0\sin(\theta'+\theta'_0)$, where $\theta'_0$ is the position angle of the rotation axis and $A_0$ corresponds to two times the amplitude of rotation modified by a factor of $\sin i$.
The angle $i$ is the unknown inclination of the GC with respect to the line of sight. Since stellar proper motions for this object are rather unreliable, there is no way to estimate $\sin i$ \citep[see for example][]{bianchini+2013} and the amplitude of rotation is just a lower limit of the true intrinsic rotation velocity for NGC~4372.
The results of the $\chi^2$-fit are $A_0=2.0\pm0.2$~\kms~($v_{rot}\sin i = 1.0\pm0.1$~\kms) and the projected rotation axis lies at a Position Angle of $136^{\circ}\pm7^{\circ}$, measured from North through East. The results remain the same if we fit the sine law to the discrete velocities of the sample at the position angle of each star, instead to the binned picture presented in Figure \ref{fig:rotation}.

\begin{figure}
   \centering
   \includegraphics[width=\hsize]{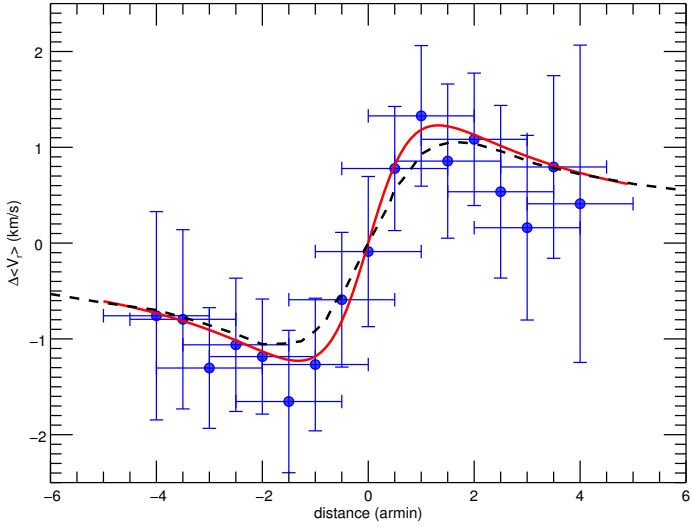}
      \caption{Rotation profile of NGC 4372, where the abscissa shows the distance from the centre of the GC along the axis perpendicular to the rotation axis and the ordinate shows the mean offset from the GC's systemic velocity in different overlapping bins. The horizontal error bars indicate the size of the selected bins, while the vertical error bars indicate the formal uncertainty of the mean velocity offset. The best fit, according to Eq. (10) is overplotted with a red line and the rotating model is shown with a black dashed line. 
              }
         \label{fig:rotation_curve}
   \end{figure}

The amplitude of rotation, however, varies with radius as naturally expected with the presence of differential rotation \citep[e.g.][]{bellazzini+2012,mackey+2013}.
The simplest way to show this for NGC~4372 is to construct a rotation profile (Figure \ref{fig:rotation_curve}). The figure shows the mean difference between the velocity measured in different overlapping bins along an axis perpendicular to the rotation axis and the systemic radial velocity of the cluster.
The resulting curve was then fitted by a simple rotation profile of the form \citep[as in][]{mackey+2013}:
\begin{equation}
 v_{rot} = \frac{2A_{rot}}{r_{peak}}\times\frac{X_{\theta'_0}}{1+(X_{\theta'_0}/r_{peak})^2}
\end{equation}
where $r_{peak}$ is the projected radius at which the maximum amplitude of rotation $A_{rot}$ is measured and $X_{\theta'_0}$ is the distance in arcmin from the cluster's centre along an axis perpendicular to the axis of rotation. For NGC~4372, we found $A_{rot} = 1.2\pm0.25$~\kms, at $r_{peak} = 1.3\pm0.5\arcmin$ from the cluster centre.

The estimated $A_{rot}/\sigma_0$ ratio is $0.26\pm0.07$. Its meaning is further discussed in Section 5.

\begin{figure}
   \centering
   \includegraphics[width=\hsize]{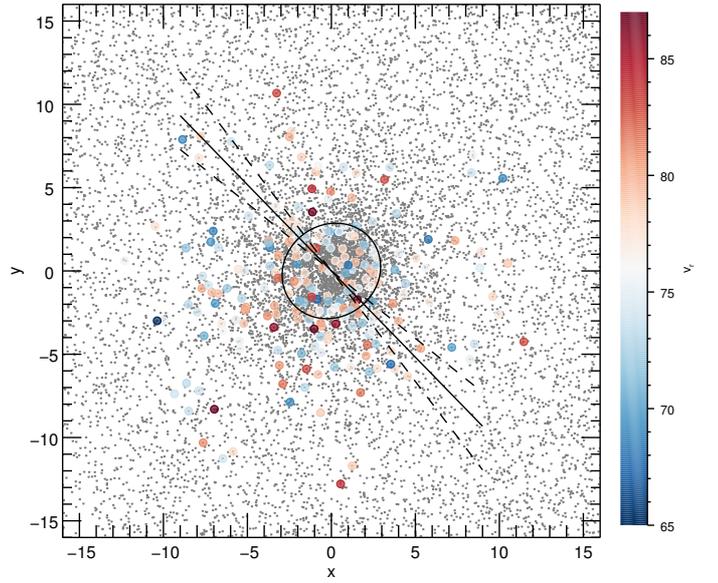}
      \caption{Spatial extent of the RGB/AGB and turn-off stars of NGC~4372. Stars, for which we have radial velocity measurements are highlighted with colourful symbols, denoting their line-of-sight velocity. The best fitting ellipsoid for this set of stars (see the last column of Table \ref{tab:mcmc_fit}) and the rotation axis are overimposed.
              }
         \label{fig:rotation_2D}
   \end{figure}

It is worth to note that the estimated angle of the major axis of the ellipsoid that best fits the number-density profile $\theta\sim40^\circ$ is perpendicular to the estimated position angle of the rotation axis of NGC~4372 ($\theta' = 136^\circ$). Since our MCMC fitting algorithm is most sensitive to the ellipticity in the inner parts of the cluster and that the maximum of rotation is found well within the half-light radius of NGC~4372, we can already conclude that systemic rotation is likely the main driver of the flattening of NGC~4372.
In Figure \ref{fig:rotation_2D} we show the spatial extent of the RGB/AGB and turn-off stars from our photometric catalogue. The stars, for which we have radial velocity measurements are highlighted with colourful symbols denoting their line-of-sight velocity. The best fitting ellipsoid for this set of stars (see the last column of Table \ref{tab:mcmc_fit}) and the rotation axis are overimposed.

\section{Discussion}

\subsection{A dynamical model for NGC~4372}

The present section is motivated by the need to provide a global and realistic dynamical interpretation of NGC 4372, taking into consideration all the morphological and kinematic properties we have collected and discussed in the previous sections. In order to carry out a complete dynamical description of NGC 4372 we compare our full set of observations with a family of physically motivated distribution-function based models, recently applied to a selected sample of GCs \citep{varri+bertin2012,bianchini+2013}.
These self-consistent models have been specifically constructed to describe quasi-relaxed stellar systems and to take into account realistic differential rotation, axisymmetry and pressure anisotropy. The models are defined by four dimensionless parameters (concentration parameter $\Psi$, rotation strength parameter $\chi$, and the parameters $\mu$ and $\xi$ determining the shape of the rotation profile). A full description of the distribution function and of the parameter space is provided in \citet{varri+bertin2012}.

The comparison between the differentially rotating models and observations requires to specify the four dimensionless parameters and five additional quantities: three physical scales (i.e., the radial scale $r_0$, the central surface density $n_0$, and the velocity scale $v_0$), the inclination angle $i$ between the rotation axis and the line-of-sight direction, and a foreground contamination term $n_f$ (to be added to the surface density profile).
For simplicity, two fixed inclination angles at $i=45^{\circ}$ and $i=90^{\circ}$ are adopted.\footnote{The inclination angle of the rotating axis is generally not known for globular clusters; given the fact we observe small but yet significant flattening and rotation, we are led to conclude that  the angle is likely to be different from $i=0$. Strictly statistically, $<\sin i>=\frac{\pi}{4}$, therefore the average inclination angle is $<i> = 52^{\circ}$. A detailed exploration of this additional parameter is beyond the goal of our study.}
The fit procedure is conducted in two steps. First, we determine the dimensionless parameters such to reproduce the observed value of $A_{rot}/\sigma_0$ and the observed position of the rotation peak (for further details see Sect. 3.1 and 3.5 of Bianchini et al. 2013). Second, we calculate the physical scales by minimizing $\chi^2$ simultaneously for the combined photometry (surface density profile) and kinematics (dispersion profile and rotation profile). This provides at once all the constraints needed to determine the best fit dynamical model.

\begin{table}
\begin{center}
\caption{Parameters derived from the rotating model.}\label{tab:model_fit}
{\small
 \begin{tabular}{cccc}
\hline
  & $V<20$~mag &  $V < 19$~mag & $V < 19$~mag \\
  & $i=45^{\circ}$ & $i=45^{\circ}$ & $i=90^{\circ}$ \\
\hline
\multicolumn{4}{c}{best fit physical scales}\\
\hline
$r_0$ (arcmin) & $2.49\pm0.05$ & $2.19\pm0.06$ & $2.23\pm0.07$ \\
$v_0$ (\kms) & $4.81\pm0.25$ & $4.98\pm0.26$ & $5.14\pm0.28$ \\
$n_0$ & $358\pm8$ & $165.2\pm6.9$ & $168.2\pm10.8$ \\
$n_f$ & $12.0\pm0.3$ & $6.7\pm0.23$ & $6.7\pm0.15$ \\
\hline
\multicolumn{4}{c}{derived quantities}\\
\hline
$r_h$ (arcmin) & $3.64\pm0.07$ & $3.20\pm0.09$ & $3.12\pm0.10$ \\
$r_t$ (arcmin) & $37.3\pm0.75$ & $32.8\pm0.9$ & $33.4\pm1.0$\\
$r_c$ (arcmin) & $2.25\pm0.05$ & $1.98\pm0.05$ & $1.94\pm0.06$ \\
$c = \log(r_t/r_c)$ & $1.22\pm0.01$ & $1.23\pm0.02$ & $1.24\pm0.03$ \\
$M_{dyn}$ ($10^5~\mathrm{M_{\odot}}$)  & $1.88\pm0.38$ & $1.97\pm0.54$ & $1.94\pm0.61$ \\
$M/L_V~\mathrm{(M_{\odot}/L_{\odot})}$ & $1.7\pm0.4$ & $1.8\pm0.5$ & $1.8\pm0.6$ \\
\hline
 \end{tabular}
\par}
\end{center}
\end{table}

In the subsequent dynamical analysis stars will be used as kinematic tracers and we will assume that the stars in our kinematic data sets trace the true stellar mass population of the system.
We recall that our dynamical models are one mass component models, and therefore can be applied when the stellar population of the system is homogeneous. 
However, in the case of the presence of mass segregation and energy equipartition we expect stars of different masses to have different spatial distribution and different kinematics \citep{trenti+vandermarel2013}.

Some additional attention is required when using simultaneously photometry and kinematics that refer to stars of different magnitude ranges.
Therefore, we compute two models, where the first takes as an input the number density distribution of all stars brighter than $V=20$~mag (Figure \ref{fig:number_density}), and the second one uses only stars brighter than $V=19$~mag (Figure \ref{fig:number_density_bright}).
A better $\chi^2$ is obtained when the number density profile of the brighter stars only is used.
The computed dimensionless parameters are as follows $\Psi=5$, $\chi=0.16$, $\mu=0.5$, $\xi=3$. The derived physical scales and structural parameters (core radius, half-light radius, concentration, total mass, $M/L_V$) are reported in Table \ref{tab:model_fit}.

The best fit model, constrained using the brighter RGB/AGB and turn-off stars is able to reproduce satisfactorily both the photometric and the kinematic radial profiles.
In particular, the model reproduces well the central region and the outer part of the number density profile (see Figure \ref{fig:number_density_bright}).
For the line-of-sight kinematic profiles, the model is able to reproduce simultaneously the shape of the velocity dispersion profile and of the rotation profile, matching the characteristic rigid rotation behavior in the central regions, the velocity peak, and the subsequent decline (see Figures \ref{fig:plummer} and \ref{fig:rotation_curve}).
We ran a 2-dimensional realization of the rotating model through the MCMC fitting routine to estimate its flattening due to rotation. Interestingly, the model does not allow significant deviations from spherical symmetry.\footnote{Complex interplay between rotation and velocity anisotropy can contribute to the final, morphological properties of the modelled stellar system.}

The derived estimates of the total dynamical mass from both models of $M_{dyn}=1.9\pm0.4\times10^5~M_{\odot}$ and $2.0\pm0.5\times10^5~M_{\odot}$ are in agreement with the virial estimate reported in Section 4.1, and suggests a mass-to-light ratio $M/L_V$ between $1.4$ and $2.3~\mathrm{M_{\odot}/L_{\odot}}$.

To assess the impact of the unknown inclination angle, we also computed a model assuming the extreme case that we see the cluster edge on ($i=90^{\circ}$). The derived parameters from this model are also reported in Table \ref{tab:model_fit}.
Although this assumption provides slightly better $\chi^2$-fit, the derived structural quantities are essentially unchanged and we can conclude that a reasonable choice of the inclination angle (between $45^{\circ}$ and $90^{\circ}$) does not have a real impact in our final results. Inclination angles below $30^{\circ}$ make the rotation signal practically undetectable.

\subsection{On rotation and ellipticity}

GCs are to a high degree spherically symmetric systems but mild deviations from the perfect sphere (ellipticities up to $0.20$) are observed in most of them.
Different reasons for what could cause the flattening are discussed in the literature. Amongst other, internal rotation, pressure anisotropy, and external tides have been suggested to have a significant impact \citep[see][]{goodwin97,gnedin+99,vandenbergh2008,bianchini+2013}.
\citet{mackey+vandenbergh2005} also suggested that the observed shape of GCs could be governed by a tri-axial, dark matter, mini halo, in which GCs hypothetically could reside.
Although the main paradigm is that there is little or no dark matter around GCs \citep{baumgardt+2005a,baumgardt+2009,lane+2009,lane+2010,sollima+2012,ibata+2013}, this subject is not yet fully examined due to the lack of radially extended kinematic data sets suitable for dynamical studies in the majority of GCs \citep{zocchi+2012}, unless tidal streams are observed \citep{mashchenko+Sills2005a,mashchenko+Sills2005b}.

\begin{figure}
   \centering
   \includegraphics[width=\hsize]{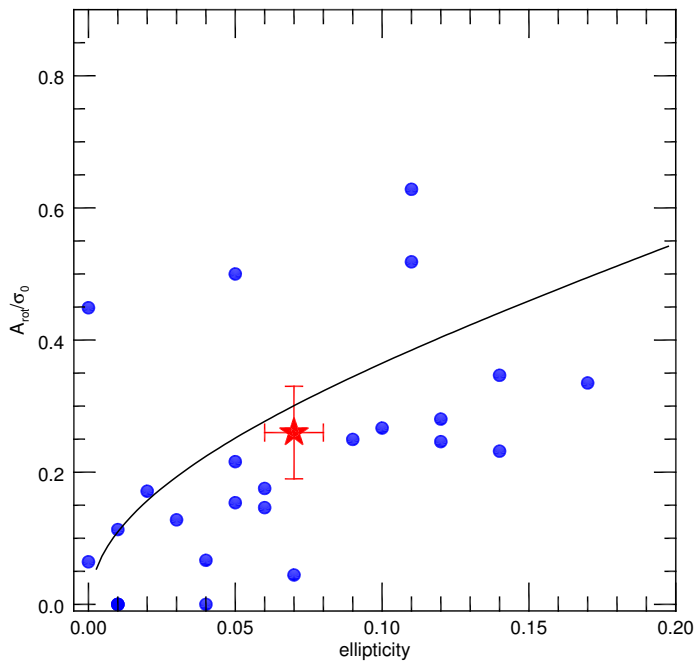}
      \caption{The rotation to velocity dispersion ratio plotted as a function of GC ellipticity. The sample of GCs with known $A_{rot}/\sigma_0$ values comes from \citet{bellazzini+2012,bianchini+2013} and the ellipticity information is taken from \citet[][2010 version]{harris96}, except for NGC~4372 (red star), for which we use our own estimates. A model of an isotropic, rotating spheroid is overimposed for comparison \citep{binney2005}. 
              }
         \label{fig:rotation_ellipticity}
   \end{figure}

Plots of the $A_{rot}/\sigma_0$ ratio vs. ellipticity (Figure \ref{fig:rotation_ellipticity}) are a common tool used to assess the importance of rotation in shaping stellar systems \citep[][in the context of elliptical galaxies]{davies+83,emsellem+2011}.
Both the rotational velocity estimates and the observed ellipticities depend on the unknown inclination angles to the line-of-sight and thus constitute  lower limits.
Taking that into account and the limited radial extent of the available radial velocity data, most of the GCs plotted in Figure \ref{fig:rotation_ellipticity} have ellipticities consistent with the flattening caused by rotation according to a model of a self-gravitating, rotating sphere \citep{binney2005}. Although the plot shows that in most cases the flattening indeed seems to be caused by significant internal rotation, this is difficult to be conclusively proven with the existing data sets.
Both rotation and ellipticity vary with the radial distance from the cluster centre \citep{geyer+83,bianchini+2013}, and different factors (including anisotropy) may have different impact at different projected radii. There is, for example, a large systematic discrepancy between the two most extensive studies on GCs' ellipticities, \citet{white+shawl87} and \citet{chen+chen2010}. They are based on different types of data and methods, which are sensitive to different radial distances.

NGC~4372 lies firmly on the sequence described by other GCs in the $A_{rot}/\sigma_0$ vs. $\epsilon$ diagram (Figure \ref{fig:rotation_ellipticity}) and in good agreement with the dynamical model. This implies that its flattening is mostly caused by its significant internal rotation. This conclusion is additionally supported by the excellent alignment of the rotational axis and the orientation of the best fitting ellipse and is somewhat surprising given its proximity to the Galactic disk, where external tides are expected to play a significant role.
The lack of proper motions prevents us to directly assess the role of anisotropy in velocity space.


\subsection{How does rotation affect other GC parameters?}

\begin{figure*}
   \centering
   \includegraphics[width=\hsize]{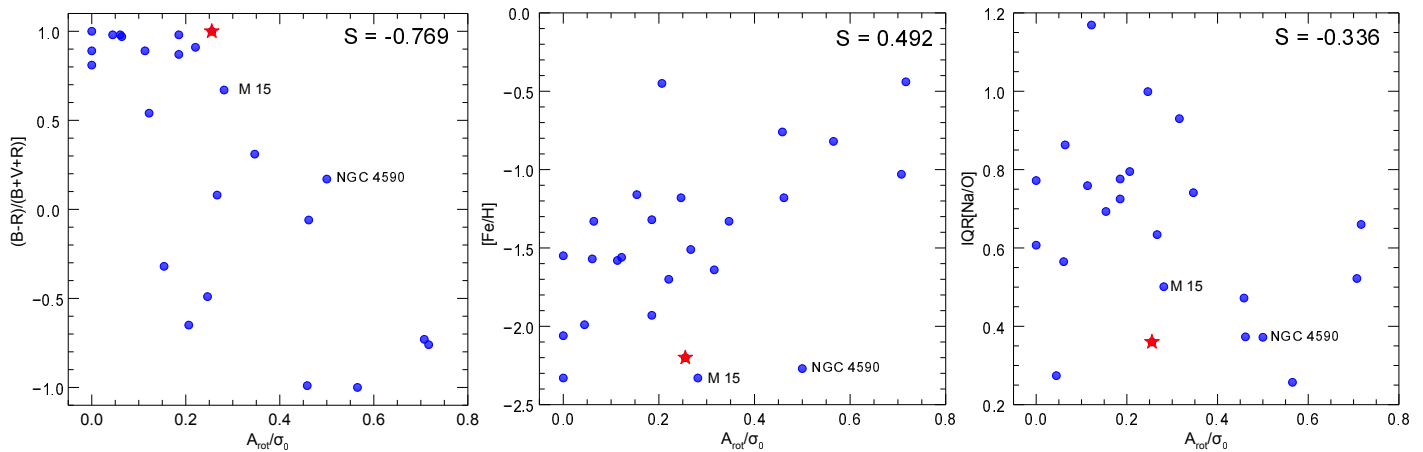}
      \caption{The $A_{rot}/\sigma_0$ ratio in a number of GCs as a function of different cluster parameters: horizontal branch morphology ({\it left panel}), [Fe/H] ({\it mid panel}), and IQR(Na/O) ({\it right panel}). The Spearman rank correlation coefficient is shown at the upper right corner of each panel. Rotation data are taken from \citet{bellazzini+2012}. NGC~4372 has a red star symbol.
              }
         \label{fig:rotation_properties}
   \end{figure*}

\citet{bellazzini+2012} explored the dependence of various cluster parameters on the $A_{rot}$ and $A_{rot}/\sigma_0$ ratio using the kinematic results for a sample of 25 Galactic GCs. They found a very strong correlation between the amount of rotation in a GC and its horizontal branch (HB) morphology, namely that GCs with blue HBs are slower rotators than those with red or extended HBs (Figure \ref{fig:rotation_properties}). NGC~4372 is not an exception and firmly takes its place as a relatively slow rotator with a very blue HB on this diagram.
The HB morphology, however, is one of the most complex parameters that characterise GCs \citep[see][for a detailed review]{catelan2009} and although not completely unexpected, such a strong correlation of the rotation with the HB morphology is difficult to explain since it has to be due to the superposition of multiple effects.
Metallicity is the first parameter that shapes the HB morphology \citep{lee+90,fusi-pecci+93,gratton+2010} and thus, it is not surprising that \citet{bellazzini+2012} also found a significant correlation between the [Fe/H] abundance and the $A_{rot}/\sigma_0$ ratio.
They noted however, that the correlation with metallicity is not as strong as the dependence on the HB morphology and it is unlikely to be the sole parameter. The Spearman rank correlation coefficient determined by \citep{bellazzini+2012} for each of the discussed relations is shown in Figure \ref{fig:rotation_properties} to give a feeling for their significance.
NGC~4372 is one of the most metal-poor GCs in the Galaxy and Figure \ref{fig:rotation_properties} shows that it has a somewhat larger amount of ordered motion for its metallicity. That brings it in line with other outliers on this diagram like NGC~7078 (M~15) and NGC~4590.

Age is widely accepted as being the second most important parameter that shapes the HB \citep[e.g.][McDonald \& Zijlstra, in prep.]{searle+zinn78,lee+94,mackey+gilmore2004,mackey+vandenbergh2005} and is likely to have a significant impact on the rotation of GCs in the sense that older clusters are expected to have dissipated their angular momentum through dynamical relaxation, or could be slowed down by tidal interactions with the Galaxy \citep{goodwin97}. In this respect, one could see the left panel of Figure \ref{fig:rotation_properties} as an ensemble of slowly rotating old halo GCs with blue HBs and their rapidly rotating counterparts of young halo clusters with extended or red HBs.

Additionally, \citet{bellazzini+2012} suggested a weak inverse relation between $A_{rot}/\sigma_0$ and the inter-quartile range of the [Na/O] abundance ratio. The latter is a proxy for the extent of the Na and O variations and thus is an indicator of the occurrence of multiple populations in GCs \citep[see][]{carretta+2009b,gratton+2012}.
While this dependence is not very significant (Spearman rank $-0.34$), it is likely to be a secondary effect of the HB morphology dependence, since the extended spreads of Na and O abundances in GCs are linked to the extent of He abundances in GCs stars \citep{dantona+2002}.
Helium is important for shaping the HB, as He-enriched stars reach hotter temperatures at the HB stage and GCs with pronounced multiple populations have generally more extended HBs.
If we assumed that the relation between the extent of the [Na/O] abundance ratio in a GC and its amount of internal rotation represent a genuine formation process, we would expect to observe a direct dependence instead of a reversed one. Indeed, according to the most popular scenarios of GCs formation \citep{dercole+2008,decressin+2007,bekki2010}, GCs form a second, dynamically cold, and rapidly rotating stellar population. If this were the case, we would still see the clusters with more numerous, chemically-enhanced population to be faster rotators.
NGC~4372 shows the typical of GCs Na-O anticorrelation as our chemical abundance analysis shows. The value for the IQR[Na/O] shown in Figure \ref{fig:rotation_properties} is estimated from abundance measurements of the same GIRAFFE spectroscopic sample presented in this work. The full abundance analysis will be presented in a series of subsequent articles.

The amplitude of internal rotation in GCs is also connected to their absolute magnitudes and central velocity dispersions \citep[see][]{bellazzini+2012}.
Both quantities are tightly linked to the clusters total mass. NGC~4372 is not an exception in this respect.

\begin{figure}
   \centering
   \includegraphics[width=\hsize]{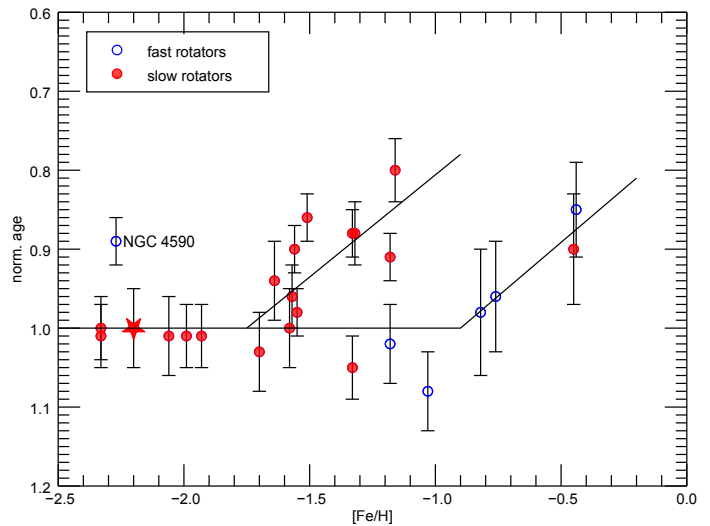}
      \caption{Age-metallicity relation for GCs with known rotation properties. The normalised ages are taken from \citet{marin-franch+2009} based on the \citet{ZW84} scale. The points are colour-coded based on the $A_{rot}/\sigma_0$ ratio. Slow rotators are marked with red and fast rotators with blue. NGC~4372 is marked with a red star. The two sequences of GCs supposedly accreted into the Milky Way halo and born in-situ are marked with black lines.
              }
         \label{fig:age_metallicity}
   \end{figure}

In the end, we explore the age-metallicity relation of GCs with respect to their rotation properties (Figure \ref{fig:age_metallicity}). Normalised ages are taken from the work of \citet{marin-franch+2009} based on the \citet{ZW84} metallicity scale. We have divided the GCs from \citet{bellazzini+2012} and NGC~4372 into slow and fast rotators according to their $A_{rot}/\sigma_0$ ratio with respect to the prediction of the isotropic, rotating model shown in Figure \ref{fig:rotation_ellipticity}. GCs lying significantly above the predictions of the model are considered as fast rotators.
The two well known branches of GCs in the age-metallicity relation are often interpreted as clusters born in-situ and accreted on a later stage in the Milky Way halo as indicated in the figure \citep{marin-franch+2009,leaman+2013}.
Figure \ref{fig:age_metallicity} shows that the majority of fast-rotating clusters are the ones supposedly born in-situ, while the slow rotators occupy the more metal poor branch associated with the accreted GCs.

\section{Summary}

We presented the first in depth study of the structure and kinematics of the old, metal-poor GC NGC~4372. We used archival deep $V$- and $I$-band images from the WFI camera mounted at the MPG/ESO 2.2 m telescope to obtain a CMD of the cluster. NGC~4372 is known to suffer from a severe differential reddening. We used a well established method to correct for it and showed that the mean $E(B-V)$ is $0.5$~mag but varies between $0.3$ and $0.8$~mag across the observed field. We made a cut at $V=20$~mag in the extinction corrected CMD and used the resulting catalogue to obtain a number density profile of this GC.
Using a MCMC maximum likelihood fitting procedure we derived the centroid of the cluster, its half-light radius ($r_h=3.44\pm0.04\arcmin$), ellipticity ($\epsilon=0.08\pm0.01$), and foreground stellar density, by considering a Plummer distribution.

In order to derive the kinematic properties of NGC~4372, we used high-resolution spectroscopic observations from the FLAMES spectrograph at the VLT. Based on precise radial velocity measurements and metallicity estimates, we selected a clean sample of 131 NGC~4372 RGB/AGB stars.
With this radial velocity sample we derived a rotation profile extending to the cluster half-light radius. These observations show that the cluster has a significant internal rotation with a maximum amplitude of $A_{rot}=1.2\pm0.25$~\kms.
On the other hand, the best fit velocity dispersion profile indicates a central velocity dispersion of $\sigma_0 = 4.56\pm0.3$~\kms. The resulting $A_{rot}/\sigma_0$~ratio of $0.26\pm0.07$ is relatively large for NGC~4372's low metallicity and old age.
Our results show that NGC~4372 is flattened in the direction of its internal rotation (i.e. perpendicular to its rotation axis), which is the most likely reason for deviations from sphericity.

Our observational results allow us to construct a realistic dynamical interpretation of NGC~4372. We compared our full set of observations with a family of physically motivated, distribution-function based models, specifically constructed to describe quasi-relaxed, differentially rotating stellar systems. The best fitting model is a good representation of the number density and velocity dispersion profile of this GC, as well as its differential rotation. However, being highly spherical, the model fails to reproduce the observed flattening of the cluster.
Based on this model, the total dynamical mass of NGC~4372 is $\sim2\times10^5~\mathrm{M_{\odot}}$ with a mass-to-light ratio $M/L_V$ between $1.4$ and $2.3~\mathrm{M_{\odot}/L_{\odot}}$. We found that the modelled quantities depend very weakly on the adopted inclination angle.


Finally, we discuss the importance of internal rotation (particularly the $A_{rot}/\sigma_0$~ratio) of NGC~4372 to its morphology and chemical composition by comparing it to similar studies of other GCs \citep[see][]{bellazzini+2012}. We argue that the presumably less relaxed young halo GCs have generally higher $A_{rot}/\sigma_0$~ratios. We also show that NGC~4372 (an archetypical old halo GC) has unusually high $A_{rot}/\sigma_0$~ratio for its low metallicity, but it could still be considered a slow rotator when compared to the young halo GC population.
When we consider the two distinct branches of GCs in the age-metallicity relation, we notice that the fast rotating GCs are the ones presumably born in-situ, while the slow rotators occupy predominantly the branch of the presumably accreted GCs.

\begin{acknowledgements}
We thank Benjamin Hendricks for valuable discussions.
NK, AK, and MJF acknowledge the Deutsche Forschungsgemeinschaft for funding from  Emmy-Noether grant  Ko 4161/1.
CIJ gratefully acknowledges support through the Clay Fellowship administered by the Smithsonian Astrophysical Observatory.
THP acknowledges support in form of a FONDECYT Regular Project Grant (No. 1121005) and from BASAL Center for Astrophysics and Associated Technologies (PFB-06).
This work was in part supported by Sonderforschungsbereich SFB 881 ``The Milky Way System'' (subproject A4) of the German Research Foundation (DFG).
This publication makes use of data products from the Two Micron All Sky Survey, which is a joint project of the University of Massachusetts and the Infrared Processing and Analysis Center/California Institute of Technology, funded by the National Aeronautics and Space Administration and the National Science Foundation. This research used the facilities of the Canadian Astronomy Data Centre operated by the National Research Council of Canada with the support of the Canadian Space Agency. This research has made use of NASA's Astrophysics Data System.
\end{acknowledgements}

\bibliographystyle{aa}
\bibliography{mybiblio_v4}


%


\end{document}